\documentclass[letterpaper]{article} % DO NOT CHANGE THIS
\usepackage{aaai2026}
\usepackage{times}  % DO NOT CHANGE THIS
\usepackage{helvet}  % DO NOT CHANGE THIS
\usepackage{courier}  % DO NOT CHANGE THIS
\usepackage[hyphens]{url}  % DO NOT CHANGE THIS
\usepackage{graphicx} % DO NOT CHANGE THIS
\usepackage{natbib}  % DO NOT CHANGE THIS AND DO NOT ADD ANY OPTIONS TO IT
\usepackage{caption} % DO NOT CHANGE THIS AND DO NOT ADD ANY OPTIONS TO IT
\usepackage{listings}
\DeclareCaptionStyle{ruled}{labelfont=normalfont,labelsep=colon,strut=off}
\usepackage{textcomp}
\usepackage[utf8]{inputenc}
\usepackage{amsmath, amsthm, amssymb}
\usepackage{booktabs}
\usepackage{underscore}
\usepackage{algorithm}
\usepackage{algorithmic}
\usepackage{babel}
\providecommand{\remarkname}{Remark}
\providecommand{\theoremname}{Theorem}

\theoremstyle{plain}

\theoremstyle{remark}

\title{BeeRNA: tertiary structure-based RNA inverse folding \\ using Artificial Bee Colony}

\author{
Mehyar Mlaweh\textsuperscript{\rm 1},
Tristan Cazenave\textsuperscript{\rm 1},
Ines Alaya\textsuperscript{\rm 1}
}

\affiliations{
\textsuperscript{\rm 1}LAMSADE, Université Paris Dauphine -- PSL, Paris, France\\
mehyar.mlaweh@dauphine.eu, tristan.cazenave@lamsade.dauphine.fr, ialaya@parisnanterre.fr
}

\begin{document}

\maketitle

\begin{abstract}
The Ribonucleic Acid (RNA) inverse folding problem, designing nucleotide sequences that fold into specific tertiary structures, is a fundamental computational biology problem with important applications in synthetic biology and bioengineering. The design of complex three-dimensional RNA architectures remains computationally demanding and mostly unresolved, as most existing approaches focus on secondary structures. In order to address tertiary RNA inverse folding, we present BeeRNA, a bio-inspired method that employs the Artificial Bee Colony (ABC) optimization algorithm. Our approach combines base-pair distance filtering with RMSD-based structural assessment using RhoFold for structure prediction, resulting in a two-stage fitness evaluation strategy. To guarantee biologically plausible sequences with balanced GC content, the algorithm takes thermodynamic constraints and adaptive mutation rates into consideration. In this work, we focus primarily on short and medium-length RNAs ($<$ 100 nucleotides), a biologically significant regime that includes microRNAs (miRNAs), aptamers, and ribozymes, where BeeRNA achieves high structural fidelity with practical CPU runtimes. The lightweight, training-free implementation will be publicly released for reproducibility, offering a promising bio-inspired approach for RNA design in therapeutics and biotechnology.

\end{abstract}


\section{Introduction}

The Ribonucleic Acid (RNA) design problem, also known as \textit{inverse folding}, aims to identify sequences that fold into a given target structure~\cite{zadeh2011nupack}. Hofacker et al.~\cite{hofacker1994fast} first formalized the problem for secondary structures using the ViennaRNA package. The problem was later identified as NP-hard because of the large sequence space and intricate constraints such as thermodynamic stability and biological functionality~\cite{lyngso1999fast}.

RNA design helps scientists create synthetic RNA molecules for specific purposes, such as ribozymes~\cite{yamagami2019design}, microRNAs~\cite{schwab2006highly}, aptamers~\cite{hamada2018insilico}, and riboswitches~\cite{bauer2006engineered, findeiss2017design}. Yamagami et al. (2019) used computational-experimental methods to increase the catalytic efficiency of dual-pseudoknotted ribozymes~\cite{yamagami2019design}. MicroRNAs were created by Schwab et al. (2006) to silence plant genes, and synthetic riboswitches are used as biosensors in synthetic biology~\cite{findeiss2017design}.

Karaboga proposed the Artificial Bee Colony (ABC) method in~\cite{karaboga2005abc}, which models bee foraging behavior as an optimization tool in various disciplines including engineering, scheduling, and bioinformatics. The ABC algorithm imitates bee collaboration to discover optimal food sources, thus maintaining a useful balance between exploration and exploitation in complex solution spaces. Research in bioinformatics has demonstrated that ABC improves free energy calculations for lattice-based protein structures in protein inverse folding~\cite{abcprotein}. Following this, a balance-evolution ABC algorithm was proposed in 2015 to handle protein structure optimization using a three-dimensional AB off-lattice model, offering a more realistic representation of protein folding and enhanced convergence capabilities~\cite{LI20151}. However, the application of ABC to RNA inverse folding, particularly for tertiary structure prediction, remains underexplored and presents a promising direction for future research.

In this study, we introduce BeeRNA, a novel RNA inverse folding method that adapts the ABC algorithm to design sequences for target tertiary structures. BeeRNA contrasts with existing methods (e.g., ViennaRNA~\cite{lorenz2011viennarna}, NUPACK~\cite{zadeh2011nupack}) because it handles tertiary structures with ABC-based optimization. Using RhoFold~\cite{rhofold} to fold designed sequences, BeeRNA achieves better structural correctness than gRNAde~\cite{joshi2025grnade} by obtaining a 2.50~\AA\ RMSD score (Root Mean Square Deviation) on the RNASolo dataset~\cite{rnasolo}, 12.02~\AA\ on a benchmark of 14 diverse RNA structures~\cite{DasDataset}, and 14.98~\AA\ on the RFAM dataset~\cite{griffiths2003rfam2003,kalvari2018rfam13}, while gRNAde achieves 9.33~\AA, 14.63~\AA, and 16.24~\AA\ respectively. BeeRNA also achieves superior accuracy through Global Distance Test Total Score (GDT-TS) evaluation, which demonstrates its structural stability. 
Importantly, BeeRNA is specifically optimized for short and medium-length RNAs ($<$100 nucleotides)—a biologically relevant regime that includes microRNAs (miRNAs), aptamers, and ribozymes—where its training-free, metaheuristic optimization achieves a strong balance between structural precision and computational efficiency.

The paper is organized as follows: \textit{Related Work} reviews prior RNA inverse folding research; \textit{Methodology} details the BeeRNA algorithm; \textit{Results} compares its performance with existing methods; \textit{Conclusion} summarizes key findings.

\section{Related Work}
\label{previouswork}

Inverse folding, or inverse design of RNA sequences to fold into a target 3D structure, holds promise for applications in mRNA vaccines, aptamer therapy, and riboswitches. Inverse folding would deliver nucleotide sequences that take on a pre-determined tertiary structure, and it is a challenging problem considering the vast chemical space and requirements like stability and biological activity. The goal of RNA design is to identify a sequence that folds into a given tertiary structure. This section provides a concise review of the major RNA inverse folding methods rooted in deterministic, stochastic, and deep learning approaches, with a focus on the contribution of bio-inspired metaheuristics, the ABC algorithm, upon which our suggested algorithm is founded.

\subsection{Deterministic Methods}
\label{ssec:deterministic}

Deterministic methods rely on thermodynamic models to predict RNA sequences capable of folding into a desired RNA structure. ViennaRNA~\cite{lorenz2011viennarna} operates through dynamic programming to determine minimum free energy secondary structures, which allows users to develop sequences that naturally fold into target \textbf{secondary structures}. Recent advancements have explored thermodynamic integration with machine learning, as seen in methods like MXfold2~\cite{Sato2021}, which enhance secondary structure prediction by combining free energy parameters with neural networks. However, these methods face limitations because they depend on thermodynamic approximations, which prevent them from analyzing complex \textbf{tertiary structures} with 3D loops and non-local interactions. ViennaRNA only operates for secondary structure predictions; hence it fails to perform the tertiary inverse folding functions needed for this study. Deterministic methods do not have the capability to explore various conformational spaces, thus requiring the adoption of adaptive methods to analyze biologically significant tertiary structures.

\subsection{Stochastic Methods}
\label{ssec:stochastic}
Stochastic methods are ideal for exploring the vast and complex space of possible RNA sequences, as they employ probabilistic techniques to sample candidate nucleotide combinations. This flexibility enables them to model complex tertiary structures at the possible cost of slower convergence and sensitivity to hyperparameters.

A prominent example is Monte Carlo Inverse RNA Folding~\cite{cazenave2024monte}, which generates and evaluates candidate sequences using Monte Carlo simulations, then iteratively changes them to fit a target structure. Although effective at exploring nonlinear search spaces, its computational cost and parameter dependence can restrict effectiveness. Simulated annealing, inspired by metallurgical cooling processes, starts with broad exploration at a high \textbf{temperature} and gradually narrows to optimize free energy. Studies show it achieves ~80\% structural fidelity for small secondary structures ($<$ 50 nucleotides)~\cite{kai2019efficient}, but its performance reduces for tertiary folding owing to complex 3D interactions. Bio-inspired stochastic methods such as BE-ABC~\cite{LI20151} have been employed to optimize tertiary structures using a 3D off-lattice model, providing insightful analysis that can be applied for RNA tertiary inverse folding, efficiently balancing exploration and exploitation.

\subsection{Deep Learning Methods}
Deep learning approaches have transformed RNA inverse folding by learning to map structural features, such as inter-atomic distances and 3D spatial coordinates of atoms, to corresponding nucleotide sequences. 
\textbf{R3Design}~\cite{tan2025r3design} achieves 43\% native sequence recovery for simple tertiary structures by using a graph-based encoder to capture 3D RNA coordinates and an iterative decoder directed by secondary structure constraints. By simulating primary, secondary, and spatial nucleotide interactions, relational graph neural networks improve design even more. The geometric neural network (GNN)-based technique \textbf{gRNAde}~\cite{joshi2025grnade} encodes 3D conformational ensembles as graphs and employs autoregressive decoding to produce compatible sequences. It achieves fast inference ($<$1 s) and a 56.80\% native sequence recovery rate significantly outperforming Rosetta (45\%) and R3Design (43\%), with experimental success in 50\% of pseudoknot cases. But it needs a lot of pre-training on different 3D structures. \textbf{RiboDiffusion}~\cite{RiboDiffusion}, a generative diffusion model, leverages a two-module architecture combining a GVP-GNN-based structure module and a Transformer-based sequence module to iteratively refine random sequences for RNA inverse folding based on tertiary structures. It achieves an impressive 58.96\% recovery rate on sequence similarity splits (Seq. 0.8) and 66.40\% on structure similarity splits (Struct. 0.6)\footnote{Seq. 0.8 and Struct. 0.6 denote test sets clustered by 80\% sequence or 60\% structural similarity to the training set.}, outperforming gRNAde by approximately 2.16\% and 9.60\% respectively, while also demonstrating robust performance across various RNA lengths and types, including tRNA, rRNA, and ribozymes. Notably, the RiboDiffusion paper highlights that short RNAs ($<$ 50 nt) present a challenge for the model to recover the original sequence due to their flexible conformation, which impacts recovery rates and in-silico folding performance.
A recent advance, \textbf{RISoTTo}~\cite{RISoTTo}, employs a geometric transformer with a specialized diffusion process that conditions both on secondary and tertiary contact maps. It achieves a 62\% native sequence recovery rate without data splits based on sequence or structural similarity, outperforming prior models such as gRNAde and RiboDiffusion on unbiased benchmarks.
These methods excel in speed and precision but demand large datasets and struggle with short and medium RNAs.

\subsection{Bio-Inspired Metaheuristics}
\label{ssec:biometahe}

Bio-inspired metaheuristics draw from natural processes to optimize complex problems like RNA inverse folding. These algorithms steer clear of local optima by striking a balance between exploring novel solutions and taking advantage of existing ones. Some notable examples are the ABC algorithm~\cite{karaboga2005abc}, which models foraging behavior. Ant colony optimization~\cite{dorigo1996ant}, based on pheromone trails. Particle swarm optimization~\cite{kennedy1995particle}, inspired by flock behavior. And genetic algorithms~\cite{holland1975adaptation}, which mimic Darwinian evolution.

Artificial Bee Colony has demonstrated great promise for protein inverse folding. Chen et al.~\cite{abcprotein} used an effective ABC algorithm to predict protein structures on 2D and 3D lattice models, achieving lower free energy conformations (e.g., 20-amino-acid sequences with energy values improved by up to 15\% over prior methods). Their method demonstrated ABC's capacity to traverse intricate energy landscapes by optimizing hydrophobic-polar (HP) lattice models. Recent work has also explored multiobjective evolutionary computation, such as eM2dRNAs~\cite{RUBIOLARGO2023110779}, which decomposes target structures to enhance RNA sequence design, suggesting potential adaptations for tertiary folding.

While the ABC algorithm has shown success in protein inverse folding, its application to RNA inverse folding, particularly for designing sequences that fold into specific tertiary structures, remains underexplored.

\section{BeeRNA method}
\label{sec:Methodology}
\begin{figure*}[t]
    \centering
    \includegraphics[width=0.70\linewidth]{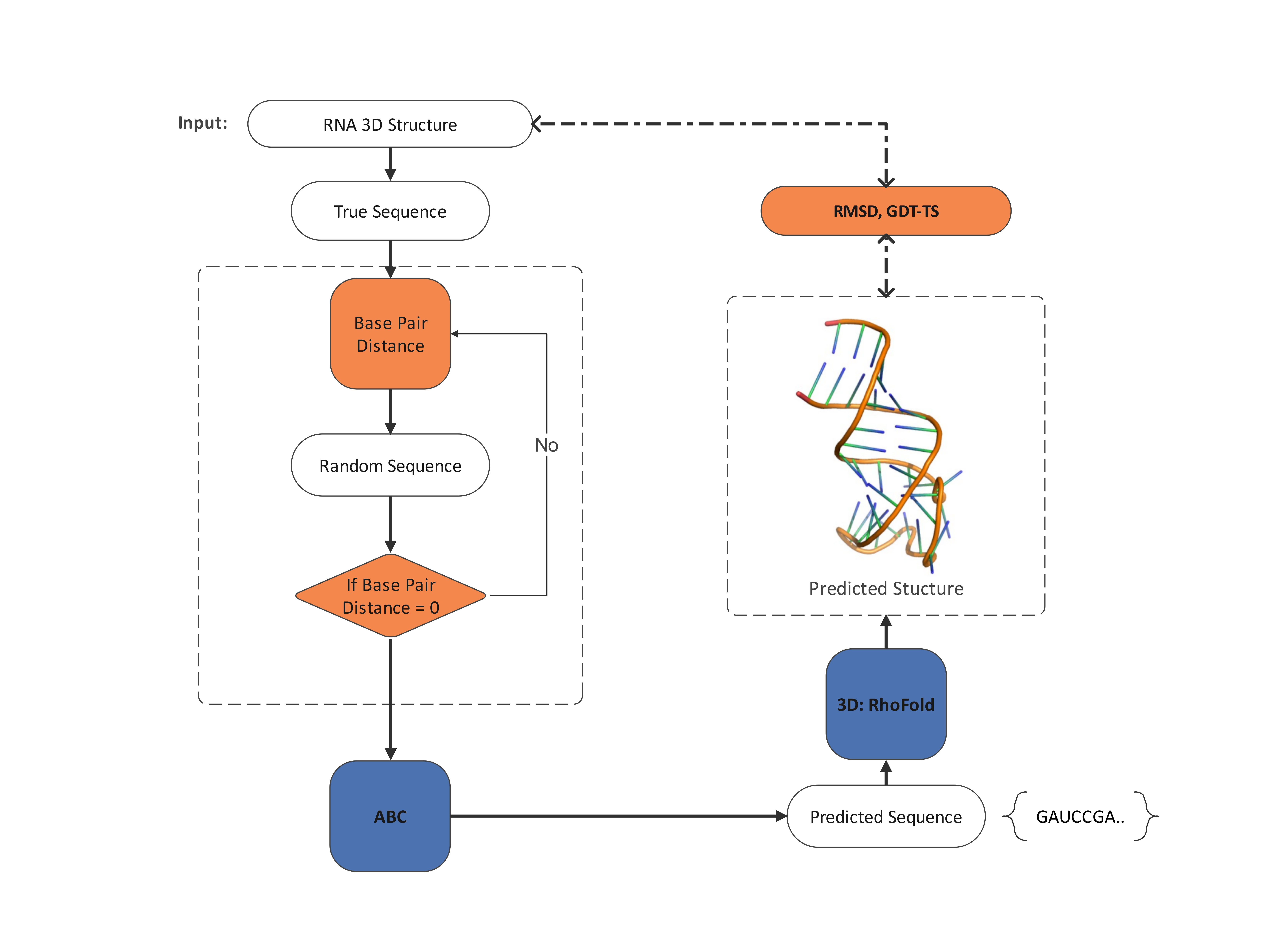}
    \caption{BeeRNA optimization process for a sample RNA structure. The flowchart illustrates the step-by-step progression of BeeRNA, starting from the initial RNA 3D structure to the final predicted sequence, using the ABC that employs RMSD calculation as the fitness function, comparing the predicted structure generated by RhoFold with the initial structure. A preliminary filter of candidate sequences is applied using base-pair distance to enhance efficiency.}
    \label{fig:methodology}
\end{figure*}

This section introduces BeeRNA, a new RNA inverse folding process that combines the ABC algorithm ~\cite{karaboga2005abc} with sequence optimization for creating target tertiary RNA structures. As seen in Figure~\ref{fig:methodology}, BeeRNA uses RhoFold~\cite{rhofold} as the structure prediction tool to evaluate the quality of the generated sequences. Notably, other recent methods such as gRNAde~\cite{joshi2025grnade} and RiboDiffusion~\cite{RiboDiffusion} also employ RhoFold to predict the tertiary structures of their designed sequences, establishing it as a common benchmark for fair structural evaluation. The AlphaFold 3 framework also shows strong potential for accurate RNA and RNA–protein structure prediction, offering future opportunities for integration.

\subsection{Problem Formulation}
The RNA inverse folding problem aims to identify a nucleotide sequence 
\( S = \{s_1, s_2, \ldots, s_n\} \), where \( s_i \in \{A, U, G, C\} \) 
and \( n \) is the sequence length, that folds into a target tertiary structure 
\( T_{\text{3D}} \), defined by 3D atomic coordinates from a PDB file. 
The objective is to find:
\[
S^* = \arg\min_S \text{RMSD}(F(S), T_{\text{3D}}),
\]
where \( F(S) \) is the 3D structure predicted by RhoFold for sequence \( S \), 
and RMSD measures the structural deviation between \( F(S) \) and \( T_{\text{3D}} \). 
Additional constraints ensure thermodynamic stability, computed using 
ViennaRNA’s minimum free energy calculations~\citep{lorenz2011viennarna}, 
and biological plausibility, such as maintaining GC content between 
40\% and 60\%~\citep{zadeh2011nupack}. 
The GDT-TS serves as a secondary metric during evaluation to assess structural similarity.

\subsection{Optimization Using the ABC Algorithm}
BeeRNA optimizes RNA sequences by integrating a two-stage fitness evaluation 
with the Artificial Bee Colony algorithm, inspired by bee foraging behavior, 
to efficiently explore the vast sequence space and identify sequences that closely 
match the target tertiary structure. The fitness function is defined as follows:
\[
\text{Fitness}(S) = 
\begin{cases} 
\infty & \text{if } \text{BPD}(S, T_{\text{sec}}) > 0, \\
\text{RMSD}(F(S), T_{\text{3D}}) & \text{if } \text{BPD}(S, T_{\text{sec}}) = 0
\end{cases}
\]
where \( S \) is the candidate sequence, 
\( T_{\text{sec}} \) is the target \textit{secondary} structure 
extracted from the tertiary reference \( T_{\text{3D}} \), 
and \(\text{BPD}(S, T_{\text{sec}})\) is the base-pair distance between 
the predicted secondary structure of \( S \) (via ViennaRNA~\cite{lorenz2011viennarna}) 
and \( T_{\text{sec}} \). 
Because ViennaRNA’s dot–bracket representation encodes only canonical and wobble pairs, 
a perfect base-pair match (\(\text{BPD} = 0\)) is achievable only when the target structure 
itself contains exclusively canonical pairs. 
For targets that include wobble or non-canonical interactions 
(as defined by the Leontis–Westhof classification), 
\(\text{BPD}\) values remain nonzero even for biologically consistent folds. 
Such small deviations are tolerated within BeeRNA’s filtering stage, 
as they typically reflect permissible tertiary motifs rather than structural errors. 
In the second stage of the fitness evaluation, \( F(S) \) is the predicted tertiary structure of \( S \) (via RhoFold~\cite{rhofold}), 
and \(\text{RMSD}(F(S), T_{\text{3D}})\) is the Root Mean Square Deviation between 
\( F(S) \) and the target tertiary structure.

RMSD is calculated as:
\[
\text{RMSD} = \sqrt{\frac{1}{N} \sum_{i=1}^N \|\mathbf{r}_i - \mathbf{r}_i'\|^2},
\]
where \( N \) is the number of aligned atoms, \( \mathbf{r}_i \) are the coordinates of the \( i \)-th atom in the predicted structure, and \( \mathbf{r}_i' \) are the corresponding coordinates in the target structure after optimal superposition. For superposition, we used \textbf{US-align}~\cite{Zhang2022.04.18.488565}, which applies the Kabsch least-squares algorithm to minimize atomic deviations. The RMSD computation considers the backbone phosphorus atom (\textbf{P}), the sugar carbon atom at position 4 (\textbf{$C4^\prime$}), and the base nitrogen atoms involved in pairing (\textbf{N1} for pyrimidines and \textbf{N9} for purines). Lower RMSD values indicate better structural alignment.

\subsubsection{Initialization:}
BeeRNA initializes a population of 40 RNA sequences (\( N = 40 \)), each with length \( n \) matching the residue count of the target sequence extracted from the target structure. Initial sequences are generated by extracting base-pairing constraints from the target secondary structure using ViennaRNA’s \textbf{fold} function~\citep{lorenz2011viennarna}. Unpaired positions (dots in the secondary structure) are assigned random nucleotides from \( \{A, U, G, C\} \), while paired positions (parentheses) receive complementary base pairs (G--C or C--G) randomly to ensure \textbf{Watson--Crick pairing}, which stabilizes RNA secondary structures~\citep{watson1953molecular}. To help ensure thermodynamic stability, the G/C content of each sequence is maintained between \textbf{40\%} and \textbf{60\%}~\citep{zadeh2011nupack}.

\subsubsection{Employed Bees Phase:}
In this phase, each of the 40 sequences in the population generates a neighbor solution by applying mutations. Mutations are performed with an adaptive mutation rate, initially set to 0.095 and adjusted based on the best RMSD achieved:
\[
\text{mutation\_rate} = \max\left(0.1,\ 0.095 \cdot e^{-\frac{\text{best\_RMSD}}{5n}}\right)
\]
where \( n \) is the sequence length. The number of mutations is \( \max(1, \lfloor \text{mutation\_rate} \cdot n \rfloor) \), applied at randomly selected positions. The nucleotide swapping mechanism between \{A, U\} and \{G, C\} at unpaired positions occurs with a 50\% chance of maintaining GC content when it surpasses 50\%. The swap mutation method involves exchanging nucleotides between two positions that exist within three positions of each other with a 20\% chance. The fitness evaluation of neighbor solutions follows a two-stage process which first checks base-pair distance before calculating RMSD when the base-pair distance is zero. When the neighbor sequence shows better fitness through lower RMSD, it replaces the original sequence; otherwise, the trial counter increments. The adaptive mutation schedule is inspired by simulated annealing, allowing BeeRNA to explore widely at early stages and gradually exploit promising regions as convergence improves. The parameters (initial rate 0.095, decay factor \(5n\), and lower bound 0.1) were determined through preliminary tuning to ensure stable convergence across diverse RNA lengths while preventing both premature stagnation and excessive randomization.

\subsubsection{Onlooker Bees Phase:}
Onlooker bees probabilistically select sequences for further exploration. The selection probability for a sequence with RMSD \( r_i \) is:
\[
p_i = \frac{e^{-r_i / \tau}}{\sum_{j=1}^N e^{-r_j / \tau}}
\]

where \( \tau = 5.0 \cdot (1 + t/T) \) is a temperature parameter that increases with iteration \( t \) (out of total iterations \( T = 40 \)), encouraging exploration early and exploitation later. Selected sequences undergo the same mutation process as in the employed bees phase, and their fitness is evaluated. If a neighbor solution improves the RMSD, it replaces the original sequence, and the trial counter resets; otherwise, the trial counter increments.

\subsubsection{Scout Bees Phase:}
Scout bees address sequences that fail to improve after 5 trials by replacing them with new randomly initialized sequences, generated as in the initialization phase. This ensures the algorithm escapes local optima and maintains diversity in the population.

\section{Experimental Results}
\label{sec:Results}
In this section, we evaluate the performance of BeeRNA for RNA inverse folding through three distinct experiments using diverse RNA datasets. These experiments assess BeeRNA’s ability to design sequences that fold into target tertiary structures, using RhoFold~\cite{rhofold} for structure prediction and RMSD and GDT-TS as metrics for structural accuracy. BeeRNA was compared against gRNAde~\cite{joshi2025grnade}, a recently introduced state-of-the-art deep learning-based method for 3D RNA design. The performance evaluation shows that gRNAde exceeds the capabilities of Rosetta~\cite{DasDataset} and RDesign~\cite{tan2025r3design} as well as FARNA (Rosetta's predecessor) across various testing scenarios that measure native sequence recovery. The demonstration of competitive or superior performance compared to gRNAde establishes BeeRNA's effectiveness and shows its ability to outperform the previous generation of RNA design methods which gRNAde improved upon. The evaluation process used fixed temperature settings of \textbf{0.1} and \textbf{16} sequence samples per structure to maintain consistency with the original gRNAde evaluation approach.
We also compare BeeRNA with RiboDiffusion~\cite{RiboDiffusion}, a diffusion-based generative method for RNA design. For a fair comparison, we adopted the same hyperparameters used in the RiboDiffusion paper: the number of sampling steps set to \textbf{50}, and \textbf{8} sequences sampled per structure. However, due to the lack of transparency about the exact training set used for RiboDiffusion constructed from the Protein Data Bank (PDB) but not publicly released, we cannot determine whether the RNASolo or Rfam datasets used in our evaluation overlap with its training data. To avoid any risk of unfair comparison, we limit our benchmarking of RiboDiffusion to the 14 designated benchmark RNAs and exclude RNASolo and Rfam examples, which are likely part of its training set.
\subsection{Experimental Setting}

\subsubsection{Datasets.}
The first experiment employed RNA sequences ($<$ 30 nucleotides) from the \textbf{RNASolo} dataset~\cite{rnasolo}, providing a curated set of short RNA structures. The second experiment focused on non-coding RNA (ncRNA)\footnote{Unlike coding RNA (e.g., mRNA) which is translated into proteins, non-coding RNA (ncRNA) functions without translation, playing structural or regulatory roles (e.g., rRNA, tRNA).} sequences from the \textbf{RFAM} database~\cite{griffiths2003rfam2003,kalvari2018rfam13}, with well-distributed lengths ranging from 25 to 200 nucleotides. Since RFAM provides secondary structures in FASTA format (available at \url{https://rfam.org/}), we used RhoFold~\cite{rhofold} to generate 3D structures in PDB format, resulting in a dataset of 88 ncRNA structures. The third experiment included a benchmark set of \textbf{14 PDB structures} from~\cite{DasDataset}, selected for their structural diversity and widespread use in RNA Design studies.

\subsubsection{Implementation Details.}
BeeRNA runs for \textbf{40} iterations with a population size of \textbf{40} sequences. These hyperparameters were selected through a small grid search to balance runtime and optimization accuracy. Larger configurations provided only marginal RMSD improvements while nearly doubling runtime due to the computational cost of RhoFold predictions performed on CPU, whereas smaller populations reduced search diversity and led to premature convergence. The chosen setting thus offers an effective trade-off between exploration depth and computational efficiency. RhoFold predictions, used in the optimization process, are performed on a CPU (due to resource constraints). The final output includes the best sequence, its RMSD, and the inference time. Predicted tertiary structures are saved as PDB files.

\subsubsection{Evaluation Metrics.}
Unlike prior methods~\cite{joshi2025grnade,tan2025r3design,RiboDiffusion,RISoTTo} that primarily assess native sequence recovery (i.e., the percentage of matching nucleotides), BeeRNA evaluates predicted sequences by comparing their folded 3D structures to the target. This structure-based evaluation is more relevant to inverse folding, as high sequence identity does not ensure structural accuracy.

\begin{figure}[t]
    \centering
    \includegraphics[width=0.30\textwidth]{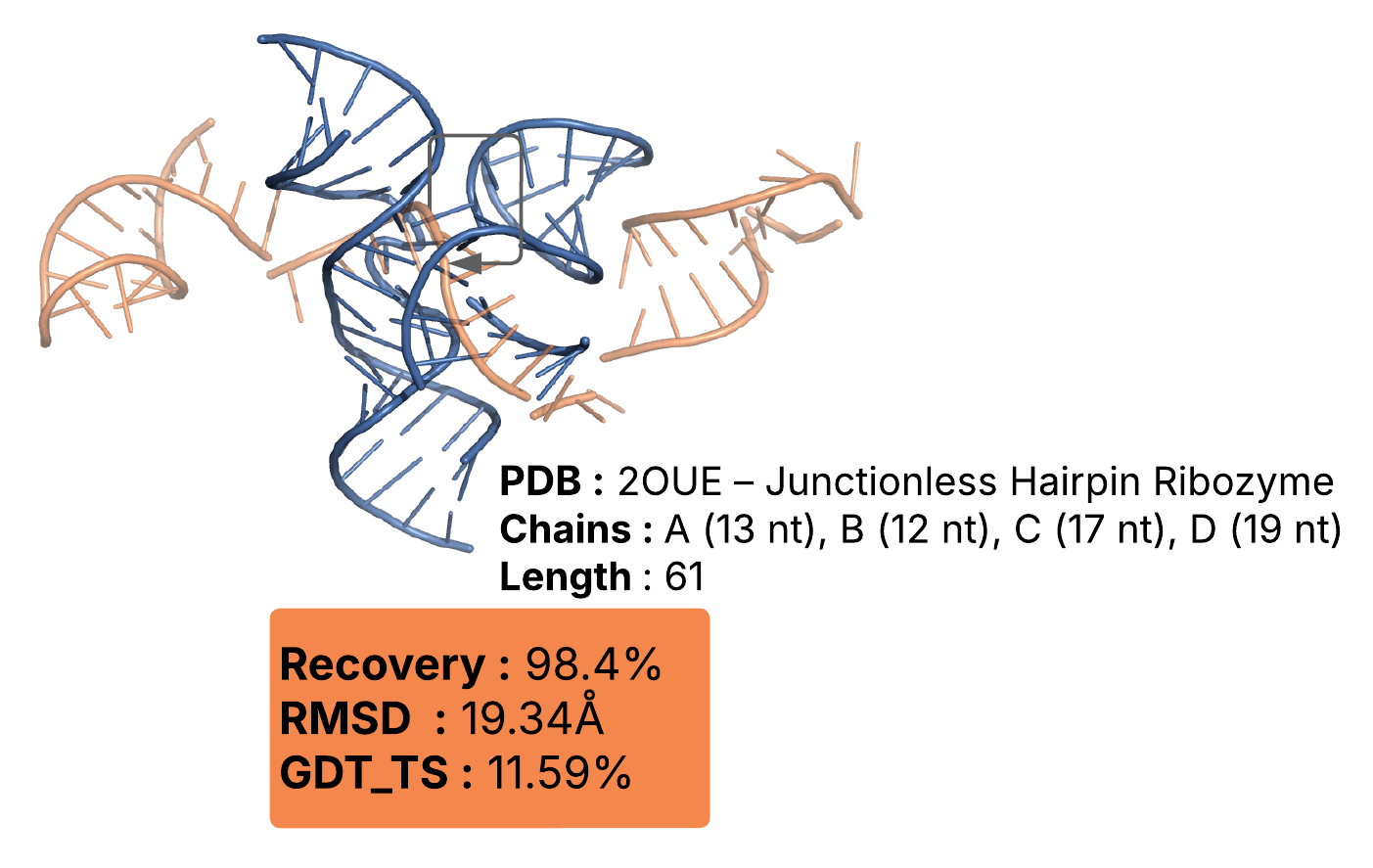}
    \caption{Predicted structure (orange) vs. sequence structure (blue) for 2OUE RNA. Despite 98.4\% native recovery, RMSD and GDT-TS show a poor structural match.}
    \label{fig:rmsd_recovery_gap}
\end{figure}

Figure~\ref{fig:rmsd_recovery_gap} illustrates this difference: starting from the native sequence of RNA \textbf{2OUE}, we introduced a single-nucleotide mutation, resulting in a sequence with \textbf{98.4\%} recovery (60/61 correct). Despite this high sequence similarity, the predicted 3D structure deviates significantly from the native one, with an RMSD of \textbf{19.34~\AA} and a GDT-TS of only \textbf{11.59\%}. 
Some of this difference may come from RhoFold’s own prediction error, but even after allowing for that, the RMSD is still much higher than normal. This demonstrates the sensitivity of tertiary contacts in short RNAs, where even minor sequence perturbations can destabilize global folding.

\textbf{RMSD:} The primary evaluation metric is RMSD, used both in the fitness function and for final assessment. RMSD quantifies the structural deviation between the predicted structure \( F(S) \) (from RhoFold) and the target structure \( T \), providing a precise measure of atomic alignment after optimal superposition using US-align~\cite{Zhang2022.04.18.488565}. BeeRNA’s optimization minimizes RMSD, ensuring the designed sequences produce tertiary structures closely matching the target.

\textbf{GDT-TS:}
The Global Distance Test Total Score~\cite{zemla2003lga} serves as a secondary metric to evaluate the structural similarity, particularly for comparing BeeRNA. GDT-TS measures the fraction of residues in the predicted structure that fall within distance cutoffs (1, 2, 4, and 8 \AA) of their corresponding residues in the target structure after superposition. It is calculated as:
\[
\text{GDT-TS} = \frac{P_1 + P_2 + P_4 + P_8}{4},
\]
where \( P_d \) is the percentage of residues within \( d \) \AA. Higher GDT-TS scores indicate better structural alignment, complementing RMSD by capturing global structural similarity.

\subsubsection{Runtime and Efficiency Evaluation.}
BeeRNA was executed on a workstation equipped with a 64-core CPU and 125~GB of RAM. 
The optimization process required approximately 3~minutes for RNAs shorter than 50~nt and 7--10~minutes for sequences up to 100~nt, including RhoFold predictions. 
Occasional runtime variation occurred when unstable or biologically implausible intermediate sequences  caused RhoFold to take longer to converge.

\subsection{Results on RNASolo}

\begin{figure}[!htbp]
    \centering
    \includegraphics[width=0.8\columnwidth]{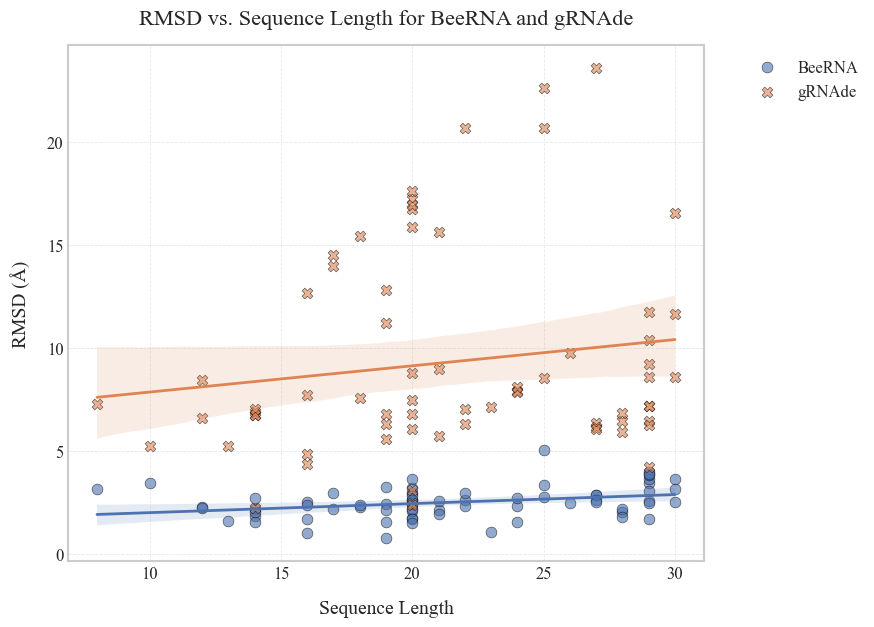}
    \caption{RMSD vs. sequence length for BeeRNA and gRNAde on RNASolo. Points represent RMSD values (\AA) for individual sequences, plotted against sequence length (3–30 nucleotides). Regression lines show RMSD trends for each method.}
    \label{fig:rmsd_01}
\end{figure}

The results demonstrate that BeeRNA significantly outperforms gRNAde on the RNASolo dataset (Table \ref{tab:avg_metrics}, Figure \ref{fig:rmsd_01}). BeeRNA achieves an average GDT-TS of 26.91\% and an average RMSD of 2.50 \AA, compared to gRNAde’s 18.97\% and 9.33 \AA, respectively, indicating superior structural accuracy and consistency. As shown in Figure \ref{fig:rmsd_01}, BeeRNA maintains stable RMSD across short sequences (3–30 nt), while gRNAde’s error increases with length. 

\begin{table}[htbp]
    \centering
    \caption{Average GDT-TS and RMSD performance metrics for the BeeRNA and gRNAde targets evaluated on the RNASolo dataset. RiboDiffusion results are excluded because its training data might include some of the RNASolo dataset}

    \begin{tabular}{lcc}
        \toprule
        \textbf{Metric} & \textbf{BeeRNA} & \textbf{gRNAde} \\
        \midrule
        RMSD (\AA) & \textbf{2.50} & 9.33 \\
        GDT-TS (\%)  & \textbf{26.91} & 18.97 \\
        \bottomrule
    \end{tabular}

    \label{tab:avg_metrics}
\end{table}

\subsection{Results on RFAM}

\begin{figure}[!htbp]
    \centering
    \includegraphics[width=0.8\columnwidth]{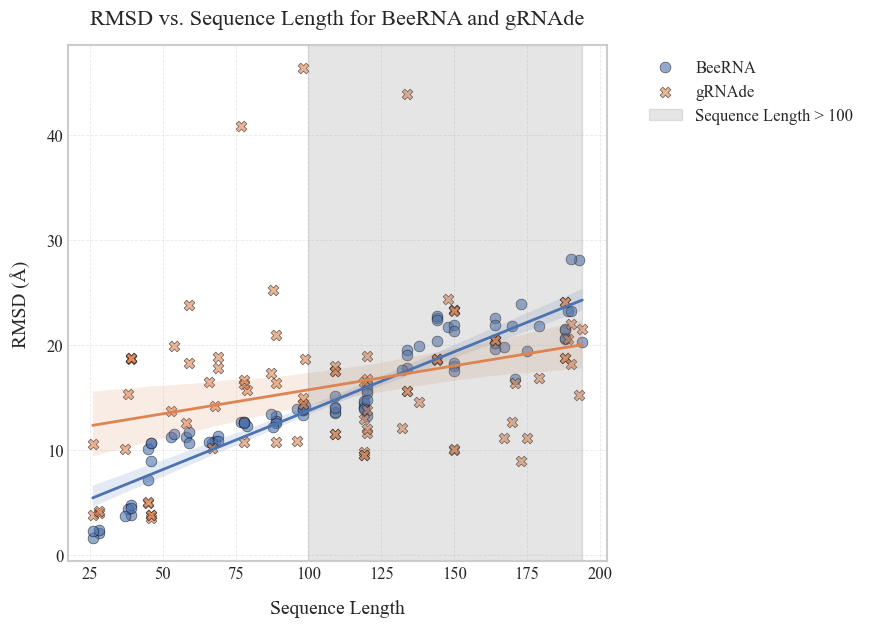}
    \caption{RMSD vs. sequence length for BeeRNA and gRNAde on RFAM. Shaded region marks sequences over 100 nucleotides}
    \label{fig:rmsd_rfam_sequence}
\end{figure}

Figure~\ref{fig:rmsd_rfam_sequence} shows the RMSD plotted against sequence length for BeeRNA and gRNAde on the RFAM dataset. Each point represents an RNA sequence, and the shaded region highlights sequences longer than 100 nucleotides. Regression lines indicate how RMSD varies with sequence length for each method.

BeeRNA achieved a higher average GDT-TS score (11.56\%) compared to gRNAde (9.77\%), and a slightly lower average RMSD (14.98\AA~vs. 16.24\AA) (Table~\ref{tab:avg_scores_rfam}). As shown in Figure \ref{fig:rmsd_rfam_sequence}, BeeRNA performs best for short and medium RNAs ($\leq$ 100 nt), maintaining lower RMSD values across this range. For longer RNAs, gRNAde slightly improves due to its deep learning architecture trained on long sequences, while BeeRNA’s stochastic, physics-driven search remains advantageous for diverse or underrepresented families. 

\begin{table}[htbp]
    \centering
    \caption{Mean GDT-TS and RMSD scores for BeeRNA and gRNAde on RFAM. RiboDiffusion results are excluded because its training data might include some of the RFAM.}
    \begin{tabular}{lcc}
        \toprule
        \textbf{Metric} & \textbf{BeeRNA} & \textbf{gRNAde} \\
        \midrule
        RMSD (\AA) & \textbf{14.98} & 16.24 \\
               GDT-TS (\%)  & \textbf{11.56} & 9.77 \\

        \bottomrule
    \end{tabular}
    \label{tab:avg_scores_rfam}
\end{table}

\subsection{Results on the 14 Benchmark RNA Structures}

\begin{table*}[htbp]
\scriptsize

\centering
\caption{Benchmark RNA structures ordered by increasing sequence length, with descriptions and RMSD (\AA) comparison between BeeRNA, gRNAde, and RiboDiffusion. Best RMSD values per structure are shown in \textbf{bold}.}
\begin{tabular}{l l c c c c}
\toprule
\textbf{PDB ID} & \textbf{Description} & \textbf{Length} $\uparrow$ & \textbf{BeeRNA} & \textbf{gRNAde} & \textbf{RiboDiffusion} \\
\midrule
1F27      & Biotin-binding RNA pseudoknot                    & 19  & \textbf{2.21}     & 14.94       & 3.47 \\
1LNT      & RNA internal loop of SRP                         & 22  & \textbf{3.69}     & 17.51       & 8.08 \\
354D      & Loop E from \textit{E. coli} 5S rRNA             & 23  & \textbf{10.68}    & 17.38       & 12.89 \\
1L2X      & Viral RNA pseudoknot                             & 27  & \textbf{3.00}              & 3.93        & 3.47 \\
1Q9A      & Sarcin/ricin domain from \textit{E. coli} 23S rRNA & 27  & \textbf{2.65}     & 6.57        & 4.35 \\
1CSL      & RRE high affinity site                           & 28  & \textbf{2.93}     & 3.36        & 12.25 \\
1ET4      & Vitamin B12 binding RNA aptamer                  & 35  & \textbf{11.34}    & 13.51       & 12.18 \\
1XPE      & ALL-RNA hairpin ribozyme                         & 46  & 20.92             & 14.26 & \textbf{10.85} \\
1X9C      & HIV-1 RNA dimerization initiation site           & 60  & \textbf{21.00}    & 25.19       & 26.79 \\
2OUE      & Junctionless hairpin ribozyme                    & 61  & \textbf{21.00}    & 22.14       & 24.71 \\
4FE5      & Guanine riboswitch aptamer                       & 67  & 11.80             & \textbf{8.42} & 9.78 \\
2GDI      & Thiamine pyrophosphate-specific riboswitch      & 78  & \textbf{8.00}     & 15.73       & 8.49 \\
2GCS      & Pre-cleavage state of glmS ribozyme             & 122 & 24.00             & 25.71 & \textbf{1.66} \\
2R8S      & \textit{Tetrahymena} ribozyme P4–P6 domain       & 159 & 26.00             & 20.20       & \textbf{5.38} \\
\bottomrule
\end{tabular}
\label{tab:rmsd_descriptions_ordered}
\end{table*}

Table~\ref{tab:rmsd_descriptions_ordered} presents the RMSD values for BeeRNA, gRNAde, and RiboDiffusion across the 14 benchmark RNA structures~\cite{DasDataset}, ordered by increasing sequence length. BeeRNA achieves the lowest RMSD values in 10 out of 14 cases, showing strong accuracy on short and medium RNAs such as 1F27 (2.21~\AA) and 1LNT (3.69~\AA). In contrast, RiboDiffusion performs best on long and complex RNAs (e.g., 2GCS and 2R8S). The higher RMSD values of 24~\AA{} and 26~\AA{} observed for these two structures result from the absence of an initial sequence achieving a BPD of zero, which constrained convergence. To handle such cases, a fixed RMSD penalty of 20~\AA{} was applied to maintain optimization continuity. Future work will refine this initialization step through Monte Carlo–based sequence generation to improve convergence on long RNA targets.

As summarized in Table~\ref{tab:avg_metrics_das} BeeRNA achieves an average RMSD of 12.02~\AA{} and GDT-TS of 15.92\%, outperforming gRNAde (14.63~\AA, 10.16\%). RiboDiffusion yields the lowest overall RMSD (10.31~\AA{}) but with greater variability across shorter targets. Overall, BeeRNA performs most reliably on RNAs below 100 nucleotides, covering most natural functional RNAs such as tRNAs, microRNAs, and aptamers, while maintaining competitive accuracy on larger structures without relying on pretraining.

\begin{table}[htbp]
\centering
\caption{Average RMSD and GDT-TS across 14 benchmark RNA structures.}
\begin{tabular}{llcc}
\toprule
\textbf{Metric} & \textbf{BeeRNA} & \textbf{gRNAde} & \textbf{RiboDiffusion}  \\
\midrule
RMSD (\AA) & 12.02 & 14.63 & 10.31 \\
GDT-TS (\%) & 15.92 & 10.16 & 22.69 \\
\bottomrule
\end{tabular}
\label{tab:avg_metrics_das}
\end{table}

The results from the RFAM and RNASolo datasets are reinforced by the benchmark evaluation, confirming BeeRNA’s strength on short and medium RNAs ($\leq$ 100 nt). This range aligns with the most functional RNAs, such as tRNAs, microRNAs, and aptamers, whose compact architectures are biologically and therapeutically relevant.
 
BeeRNA’s stochastic, ABC-based search efficiently explores these conformations without relying on pretraining, achieving low RMSD and high GDT-TS values within this length range across all datasets.

It is important to note that BeeRNA is intentionally designed for this regime. As RNA length increases, the combinatorial search space expands exponentially, making exhaustive tertiary exploration inherently more difficult for any stochastic optimizer. Consequently, slightly higher RMSD values for long RNAs reflect this inherent scaling challenge rather than a methodological limitation. In contrast, deep learning approaches like gRNAde and RiboDiffusion leverage training on long-sequence distributions to improve scalability but often struggle with the flexible and variable conformations of short RNAs.

Overall, BeeRNA achieves its objective: delivering accurate tertiary structure designs for short and medium RNAs through biologically grounded, training-free optimization. Its consistent performance across datasets highlights its adaptability and potential as a practical tool for functional RNA design within the sub-100-nt regime, while maintaining stable accuracy even as RNA length increases.

\section{Conclusion and Limitations}
\label{sec:conclusion}
In this work, we introduced BeeRNA, a novel RNA inverse folding method that adapts the ABC algorithm to design RNA sequences folding into predefined tertiary structures, effectively combining ABC exploration with biological constraints and RhoFold structural prediction to minimize structural deviations. This focus on RNAs shorter than 100 nt is biologically meaningful, as most functional RNAs, such as tRNAs, microRNAs, and aptamers, naturally fall within this range. BeeRNA outperforms gRNAde and RiboDiffusion for short and medium-length RNA sequences, with acceptable but moderately reduced performance on longer sequences due to the expanding search space, reflecting the broader challenge of scaling stochastic optimization to complex RNA folds. As BeeRNA currently uses RhoFold for tertiary structure evaluation, the method can readily integrate future RNA folding predictors as they advance, maintaining flexibility and compatibility with emerging tools. Looking ahead, the approach offers a promising foundation for hybrid strategies that combine bio-inspired search with learning-based priors and multi-objective optimization to enhance scalability, robustness, and convergence across diverse RNA lengths and structural classes.

\bibliography{beeRNA}

@article{rnasolo,
    author = {Adamczyk, Bartosz and Antczak, Maciej and Szachniuk, Marta},
    title = {RNAsolo: a repository of cleaned PDB-derived RNA 3D structures},
    journal = {Bioinformatics},
    volume = {38},
    number = {14},
    pages = {3668-3670},
    year = {2022},
    month = {06},
    issn = {1367-4803},
    doi = {10.1093/bioinformatics/btac386},
    url = {https://doi.org/10.1093/bioinformatics/btac386},
    eprint = {https://academic.oup.com/bioinformatics/article-pdf/38/14/3668/49884395/btac386.pdf},
}

@article{rhofold,
  title={Accurate RNA 3D structure prediction using a language model-based deep learning approach},
  author={Shen, Tao and Hu, Zhihang and Sun, Siqi and Liu, Di and Wong, Felix and Wang, Jiuming and Chen, Jiayang and Wang, Yixuan and Hong, Liang and Xiao, Jin and others},
  journal={Nature Methods},
  pages={1--12},
  year={2024},
  publisher={Nature Publishing Group US New York}
}

@article{DasDataset,
author = {Das, Rhiju and Karanicolas, John and Baker, David},
year = {2010},
month = {02},
pages = {291-4},
title = {Atomic accuracy in predicting and designing noncanonical RNA structure},
volume = {7},
journal = {Nature methods},
doi = {10.1038/nmeth.1433}
}

@article{lorenz2011viennarna,
  author = {Lorenz, Ronny and Bernhart, Stephan H. and Höner zu Siederdissen, Christian and Tafer, Hakim and Flamm, Christoph and Stadler, Peter F. and Hofacker, Ivo L.},
  title = {ViennaRNA Package 2.0},
  journal = {Algorithms for Molecular Biology},
  volume = {6},
  number = {1},
  pages = {26},
  year = {2011},
  url = {https://almob.biomedcentral.com/articles/10.1186/1748-7188-6-26}
}

@incollection{cazenave2024monte,
  author    = {Cazenave, Tristan and Touzani, Hamza},
  title     = {Monte Carlo Inverse RNA Folding},
  booktitle = {Methods in Molecular Biology},
  volume    = {2847},
  pages     = {205--215},
  year      = {2024},
  publisher = {Humana Press},
  doi       = {10.1007/978-1-0716-4079-1_14},
  pmid      = {39312146}
}

@article{kai2019efficient,
  author = {Kai, Zhang and Wang, Yuting and Lv, Yulin and Liu, Jun and He, Juanjuan},
  title = {An Efficient Simulated Annealing Algorithm for the RNA Secondary Structure Prediction with Pseudoknots},
  journal = {BMC Genomics},
  year = {2019},
  volume = {20},
  number = {13},
  pages = {979},
  date = {2019-12-27},
  issn = {1471-2164},
  url = {https://doi.org/10.1186/s12864-019-6300-2},
  doi = {10.1186/s12864-019-6300-2}
}

@article{LI20151,
title = {A balance-evolution artificial bee colony algorithm for protein structure optimization based on a three-dimensional AB off-lattice model},
journal = {Computational Biology and Chemistry},
volume = {54},
pages = {1-12},
year = {2015},
issn = {1476-9271},
doi = {https://doi.org/10.1016/j.compbiolchem.2014.11.004},
url = {https://www.sciencedirect.com/science/article/pii/S1476927114001467},
author = {Bai Li and Raymond Chiong and Mu Lin},

}

@article{joshi2025grnade,
  author = {Joshi, Chaitanya K. and Jamasb, Arian R. and Viñas, Ramon and Harris, Charles and Mathis, Simon V. and Morehead, Alex and Anand, Rishabh and Liò, Pietro},
  title = {gRNAde: Geometric deep learning for 3D RNA inverse design},
  journal = {Proceedings of the International Conference on Learning Representations (ICLR)},
  year = {2025},
  note = {Published as a conference paper at ICLR 2025},
  url = {https://arxiv.org/pdf/2305.14749}
}

@article{tan2025r3design,
  author = {Tan, Cheng and Zhang, Yijie and Gao, Zhangyang and Cao, Hanqun and Li, Siyuan and Ma, Siqi and Blanchette, Mathieu},
  title = {R3Design: Deep tertiary structure-based RNA sequence design and beyond},
  journal = {Briefings in Bioinformatics},
  volume = {26},
  number = {1},
  pages = {bbae682},
  year = {2025},
}

@book{holland1975adaptation,
  author = {Holland, J. H.},
  title = {Adaptation in Natural and Artificial Systems},
  publisher = {University of Michigan Press},
  year = {1975}
}

@article{RUBIOLARGO2023110779,
title = {Solving the RNA inverse folding problem through target structure decomposition and Multiobjective Evolutionary Computation},
journal = {Applied Soft Computing},
volume = {147},
pages = {110779},
year = {2023},
issn = {1568-4946},
doi = {https://doi.org/10.1016/j.asoc.2023.110779},
url = {https://www.sciencedirect.com/science/article/pii/S1568494623007974},
author = {Álvaro Rubio-Largo and Nuria Lozano-García and José M. Granado-Criado and Miguel A. Vega-Rodríguez}
}

@article{Sato2021,
  author = {Sato, Kengo and Akiyama, Manato and Sakakibara, Yasubumi},
  title = {RNA secondary structure prediction using deep learning with thermodynamic integration},
  journal = {Nature Communications},
  year = {2021},
  volume = {12},
  number = {1},
  pages = {941},
  issn = {2041-1723},
  doi = {10.1038/s41467-021-21194-4},
  url = {https://doi.org/10.1038/s41467-021-21194-4}
}

@INPROCEEDINGS{kennedy1995particle,
  author={Kennedy, J. and Eberhart, R.},
  booktitle={Proceedings of ICNN'95 - International Conference on Neural Networks}, 
  title={Particle swarm optimization}, 
  year={1995},
  volume={4},
  number={},
  pages={1942-1948 vol.4},
  doi={10.1109/ICNN.1995.488968}}

@article{karaboga2005abc,
author = {Karaboga, Dervis},
year = {2005},
month = {01},
pages = {},
title = {An Idea Based on Honey Bee Swarm for Numerical Optimization, Technical Report - TR06},
journal = {Technical Report, Erciyes University}
}

@phdthesis{dorigo1996ant,
  author = {Dorigo, M.},
  title = {The Ant Colony Optimization Metaheuristic},
  school = {Politecnico di Milano},
  year = {1996}
}

@article{abcprotein,
author = {Lin, Cheng-Jian and Su, Shih-Chieh},
year = {2012},
month = {03},
pages = {},
title = {Using an efficient artificial bee colony algorithm for protein structure prediction on lattice models},
volume = {8},
journal = {International Journal of Innovative Computing, Information and Control}
}

@article{hofacker1994fast,
  title={Fast folding and comparison of RNA secondary structures},
  author={Hofacker, Ivo L and Fontana, Walter and Stadler, Peter F and Bonhoeffer, Sebastian and Tacker, Michael and Schuster, Peter},
  journal={Monatshefte f{\"u}r Chemie / Chemical Monthly},
  volume={125},
  number={2},
  pages={167--188},
  year={1994},
  publisher={Springer},
  doi={10.1007/BF00818163}
}

@article{yamagami2019design,
  title={Design of highly active double-pseudoknotted ribozymes: a combined computational and experimental study},
  author={Yamagami, Rie and Kayedkhordeh, Mohammad and Mathews, David H and Bevilacqua, Philip C},
  journal={Nucleic Acids Research},
  volume={47},
  number={1},
  pages={29--42},
  year={2019},
  month={Jan},
  publisher={Oxford University Press},
  doi={10.1093/nar/gky1118},
  pmid={30462314},
  pmcid={PMC6326823}
}

@article{schwab2006highly,
  title={Highly specific gene silencing by artificial microRNAs in Arabidopsis},
  author={Schwab, Rebecca and Ossowski, Stephan and Riester, Markus and Warthmann, Norman and Weigel, Detlef},
  journal={The Plant Cell},
  volume={18},
  number={5},
  pages={1121--1133},
  year={2006},
  month={May},
  publisher={American Society of Plant Biologists},
  doi={10.1105/tpc.105.039834},
  pmid={16531494},
  pmcid={PMC1456875}
}

@article{hamada2018insilico,
  title={In silico approaches to RNA aptamer design},
  author={Hamada, Michiaki},
  journal={Biochimie},
  volume={145},
  pages={8--14},
  year={2018},
  month={Feb},
  publisher={Elsevier},
  doi={10.1016/j.biochi.2017.10.005},
  pmid={29032056}
}

@article{bauer2006engineered,
  title={Engineered riboswitches as novel tools in molecular biology},
  author={Bauer, Georg and Suess, Beatrix},
  journal={Journal of Biotechnology},
  volume={124},
  number={1},
  pages={4--11},
  year={2006},
  month={Jun},
  publisher={Elsevier},
  doi={10.1016/j.jbiotec.2005.12.006},
  pmid={16442180}
}

@Article{findeiss2017design,
AUTHOR = {Findeiß, Sven and Etzel, Maja and Will, Sebastian and Mörl, Mario and Stadler, Peter F.},
TITLE = {Design of Artificial Riboswitches as Biosensors},
JOURNAL = {Sensors},
VOLUME = {17},
YEAR = {2017},
NUMBER = {9},
ARTICLE-NUMBER = {1990},
URL = {https://www.mdpi.com/1424-8220/17/9/1990},
PubMedID = {28867802},
ISSN = {1424-8220},
DOI = {10.3390/s17091990}
}

@article {RISoTTo,
	author = {Bibekar, Parth and Krapp, Lucien F. and Dal Peraro, Matteo},
	title = {Context-aware geometric deep learning for RNA sequence design},
	elocation-id = {2025.06.21.660801},
	year = {2025},
	doi = {10.1101/2025.06.21.660801},
	publisher = {Cold Spring Harbor Laboratory},
	URL = {https://www.biorxiv.org/content/early/2025/06/21/2025.06.21.660801},
	eprint = {https://www.biorxiv.org/content/early/2025/06/21/2025.06.21.660801.full.pdf},
	journal = {bioRxiv}
}

@article {Zhang2022.04.18.488565,
	author = {Zhang, Chengxin and Shine, Morgan and Pyle, Anna Marie and Zhang, Yang},
	title = {US-align: Universal Structure Alignments of Proteins, Nucleic Acids, and Macromolecular Complexes},
	elocation-id = {2022.04.18.488565},
	year = {2022},
	doi = {10.1101/2022.04.18.488565},
	publisher = {Cold Spring Harbor Laboratory},
	URL = {https://www.biorxiv.org/content/early/2022/04/18/2022.04.18.488565},
	eprint = {https://www.biorxiv.org/content/early/2022/04/18/2022.04.18.488565.full.pdf},
	journal = {bioRxiv}
}

@article{griffiths2003rfam2003,
  title={Rfam: an RNA family database},
  author={Griffiths-Jones, Sam and Bateman, Alex and Marshall, Mhairi and Khanna, Ajay and Eddy, Sean R},
  journal={Nucleic Acids Research},
  volume={31},
  number={1},
  pages={439--441},
  year={2003},
  publisher={Oxford University Press}
}

@article{kalvari2018rfam13,
  title={Rfam 13.0: shifting to a genome-centric resource for non-coding RNA families},
  author={Kalvari, Ioanna and Argasinska, Joanna and Quinones-Olvera, Natalia and Nawrocki, Eric P and Rivas, Elena and Eddy, Sean R and Bateman, Alex and Finn, Robert D and Petrov, Anton I},
  journal={Nucleic Acids Research},
  volume={46},
  number={D1},
  pages={D335--D342},
  year={2018},
  publisher={Oxford University Press}
}

@article{zadeh2011nupack,
  title={NUPACK: Analysis and design of nucleic acid systems},
  author={Zadeh, Joseph N and Steenberg, Cody D and Bois, Justin S and Wolfe, Benjamin R and Pierce, Maxwell B and Khan, Ahmad R and Dirks, Robert M and Pierce, Niles A},
  journal={Journal of Computational Chemistry},
  volume={32},
  number={1},
  pages={170--173},
  year={2011},
  month={Jan},
  publisher={Wiley},
  doi={10.1002/jcc.21596},
  pmid={20645303}
}

@article{lyngso1999fast,
  title={Fast evaluation of internal loops in RNA secondary structure prediction},
  author={Lyngs{\o}, Rune B and Zuker, Michael and Pedersen, Christian N S},
  journal={Bioinformatics},
  volume={15},
  number={6},
  pages={440--445},
  year={1999},
  month={Jun},
  publisher={Oxford University Press},
  doi={10.1093/bioinformatics/15.6.440},
  pmid={10383469}
}

@article{watson1953molecular,
  title={Genetical implications of the structure of deoxyribonucleic acid},
  author={Watson, James D and Crick, Francis HC},
  journal={Nature},
  volume={171},
  number={4361},
  pages={964--967},
  year={1953},
  publisher={Nature Publishing Group}
}

@article{RiboDiffusion,
    author = {Huang, Han and Lin, Ziqian and He, Dongchen and Hong, Liang and Li, Yu},
    title = {RiboDiffusion: tertiary structure-based RNA inverse folding with generative diffusion models},
    journal = {Bioinformatics},
    volume = {40},
    number = {Supplement_1},
    pages = {i347-i356},
    year = {2024},
    month = {06},
    issn = {1367-4811}
}

@article{zemla2003lga,
  author    = {Zemla, Adam},
  title     = {LGA: A method for finding 3D similarities in protein structures},
  journal   = {Nucleic Acids Research},
  year      = {2003},
  volume    = {31},
  number    = {13},
  pages     = {3370--3374},
  doi       = {10.1093/nar/gkg571},
  pmid      = {12824330},
  pmcid     = {PMC168977}
}

\end{document}